\documentclass{emulateapj}
\usepackage{rotating}
\shorttitle{STAR FORMATION AND YOUNG POPULATION OF THE  \hii COMPLEX Sh2-294}
\shortauthors{Samal et al.}
\newcommand{\msun}{\mbox{\rm $M_{\odot}$}}

\newcommand{\lsun}{\mbox{\rm $L_{\odot}$}~}

\newcommand{\arcs}{\hbox{$^{\prime\prime}$}}

\newcommand{\arcm}{\mbox{$^{\prime}$}}

\newcommand{\av}{\mbox{$A_{\rm V}$}}

\newcommand{\hii}{\mbox{H~{\sc ii}~}}

\newcommand{\kms} {\hbox{km~s$^{-1}$~}}

\newcommand\h{H$_{2}$~}
\def\cmmt{\hbox{\kern 0.20em cm$^{-3}$}}
\begin{document}
\title{Star Formation and Young Population of the \hii Complex Sh2-294}
\author{M.R. Samal\altaffilmark{1,4},
A.K. Pandey\altaffilmark{1},
D.K. Ojha\altaffilmark{2},
N. Chauhan\altaffilmark{1},
J. Jose\altaffilmark{1}, and
B. Pandey\altaffilmark{3}
}
\email{manash.samal@oamp.fr}
\altaffiltext{1}{Aryabhatta Research Institute of Observational Sciences, Nainital 263 129, India}
\altaffiltext{2}{Tata Institute of Fundamental Research, Mumbai 400 005, India}
\altaffiltext{3}{Physics Department, D.S.B. Campus, Kumaun University, Nainital 263 129, India}
\altaffiltext{4}{Laboratoire d'Astrophysique de Marseille (UMR7326 CNRS \& Universit\'e d'Aix-Marseille), 38 rue F. Joliot-Curie, 13388 Marseille CEDEX 13}

\begin{abstract}
The Sh2-294 \hii region ionized by a single B0V star features several infrared excess sources, 
a photodissociation region, and also a group of reddened stars at its border. 
The star formation scenario in the region seems to be quite complex. 
In this paper, we present follow-up results of Sh2-294 \hii region at 3.6, 4.5, 5.8, and 8.0 $\mu$m 
observed with the {\it Spitzer Space Telescope} Infrared Array Camera
(IRAC), coupled with H$_2$ (2.12 $\mu$m) observation, to characterize the young population of the region 
and to understand its star formation history. We identified 36 young stellar object 
(YSO, Class I, Class II and Class I/II) candidates using IRAC color-color diagrams. 
It is found that Class I sources are
preferentially located at the outskirts of the \hii region and associated with
enhanced H$_2$ emission; none of them are located near the central cluster.
Combining the optical to mid-infrared (MIR) photometry of the YSO
candidates and using the spectral energy distribution fitting models, 
we constrained stellar parameters and the evolutionary status of 33 
YSO candidates. Most of them are interpreted by the model as low-mass ($<$4 \msun) YSOs; 
however, we also detected a massive YSO ($\sim$9\msun) of Class I nature, embedded
in a cloud of visual extinction of $\sim$24 mag. Present analysis suggests 
that the Class I sources are indeed younger population  of 
the region relative to Class II sources 
(age $\sim$ 4.5 $\times$ 10$^{6}$ yr). We suggest that the majority of the Class I sources, 
including the massive YSOs, are second-generation stars of the region whose
formation is possibly induced by the expansion of the \hii region powered 
by a $\sim$4 $\times$ 10$^{6}$ yr B0 main-sequence star.
\end{abstract}

\section{Introduction}
    The  massive OB stars have profound influence on 
the evolution of  molecular clouds and consequently influence  star 
formation. The stellar radiation and winds due to  massive stars in the
region can sweep  low density interstellar  matter, consequently forming a dense layer of gas
at the periphery of \hii regions or they can compress existing primordial clumps. In both the processes, the matter at
the later stage becomes unstable against self gravity to form young protostars. 
These 
two are commonly known processes that can induce new generation of star formation
around expanding \hii regions (see e.g.,  Sugitani et al. 2002; Deharveng et al. 2005; Chauhan et al. 2009; Samal et al. 2010; Ojha et al. 2011).  
Sh2-294 ($\alpha_{2000}=07^{h}16^{m}34.5^{s}$,
$\delta_{2000}=-09^{\circ}26^{\prime}38^{\prime\prime}$) is an \hii region powered by a B0V star 
and is possibly interacting  with a  molecular cloud, thus creating a photon 
dominated region (PDR) which can be  seen  towards the eastern side 
as polycyclic aromatic hydrocarbon (PAH) emission in 
MSX A-band (Samal et al. 2007; Paper-I). Using multiwavelength observations, 
Samal et al. (2007) have studied the stellar content, distribution of 
ionized, and dust emission in the region. They
identified two groups of stars, one of which is associated with the
B0V star at the center of the optically visible nebula. The second 
group (termed as ``region A'') is situated at 
the eastern side of the optical cluster and near the peak of 8 $\mu$m emission along the PDR. 
Samal et al. (2007), using the optical colour-magnitude diagram and theoretical evolutionary models,
estimated the age of the  ionizing source as 4 $\times$ 10$^{6}$ yr, whereas 
low mass pre-main-sequence (PMS) stars show an age spread of 1-5 $\times$ 10$^{6}$ yr.
On the basis of youth of the sources in  ``region A" 
(age $\leq$ 1 $\times$ 10$^{6}$ yr) compared to the age of the ionizing source ($\sim$4 $\times$ 10$^{6}$ yr), 
Samal et al. (2007) suggested Sh2-294 as a site of triggered star formation. However,  the  mechanism that is 
responsible for the initiation of new star formation in this region is still
not clearly understood; requires proper identification and characterization of
stellar sources present in the region to put a step ahead.
On the basis of high resolution near-infrared (NIR)
 and optical observations, Yun et al. (2008) estimated the age of the ionizing source of Sh2-294 as
4 $\times$ 10$^{6}$ yr,  but  suggested a higher age for the  PMS sources. 
 Due  to the absence of  spectroscopic or longer wavelength ($\lambda$ $>$ 2.2 $\mu$m) 
observations, the exact nature and evolutionary status of the PMS sources in the region could not be studied so far.
Now with high angular resolution of {\it Spitzer}  Space Telescope 
it is possible to make a more reliable membership census of the young stellar 
objects with disk and envelope of the region, as 3-8 $\mu$m bands of {\it Spitzer} reduce the degeneracy between selective interstellar extinction and intrinsic IR excess. Moreover, the nature of  YSOs can be best accomplished using their broadband SEDs and its comparison with 
more sophisticated radiative transfer SED fitting models (e.g., 
Robitaille et al. 2007). Thus, in conjugation with the available photometric 
data sets for the region, the {\it Spitzer} observations will allow us to 
characterize the individual YSOs with their SEDs and constrained 
their evolutionary status, thus will help to establish spatial  and temporal relationship among the YSOs. Therefore, we revisited the region at near- and mid-infrared windows for the follow up study of 
Sh2-294 star-forming-region (SFR), to have a better picture of
star formation activity.
The distance to Sh2-294 is uncertain and varies from 3.2 kpc to 4.8 kpc (Moffat et al. 1979; Samal et al. 2007; Yun et al. 2008). In this work, we adopted an average distance of 4.0 kpc
for our analysis and organiged the paper with the following layout. We describe the {\it Spitzer} observations and
data reduction techniques in $\S$2.
In $\S$3, the observational results are presented, which includes
morphology of the region, identification of the  YSOs and their properties obtained
with SED modelling. The star formation scenario in the region is discussed in
$\S$4 and $\S$5 summarizes our results.

\section{Observations and Data Reduction}
The archived MIR data for the region were obtained with the
Infrared Array Camera (IRAC; Fazio et al. 2004a) on board the {\it Spitzer}
Space Telescope. IRAC has four wavelength bands centered at 3.6, 4.5, 5.8, 
and 8.0 $\mu$m, each of which has a field of view of 
$\sim$5\arcm.2 $\times$ 5\arcm.2.
The pixel size in all the four bands is $\sim$ $1\farcs22$.
The Corrected Basic Calibrated Data (CBCD) images were downloaded from the {\it Spitzer}
Space Telescope Archive using the Leopard package.  The IRAC observations of the region
were taken on  2006 November 25 (Program ID 30734, AOR key 18902784: Massive Star Clusters, PI: Donald Figer).
Mosaics were built at the native instrument resolution of $1\farcs22$ per pixel with the standard BCDs
using the MOPEX (Mosaicker and Point Source Extractor) software program provided by
{\it Spitzer} Science Center. 

                    We performed aperture photometry on the IRAC images using the DAOPHOT package of IRAF, with a source detection at the 5$\sigma$ level
 above the average local background. Due to the crowded nature of the field,
an aperture radius of 2 pixels  and a sky annulus extending from 2 to 6 pixels
was used. We adopted the zero-point magnitudes for the standard aperture radius (10 pixels) and 
background annulus of (10-20 pixels) of 19.67, 18.92, 16.86, and 17.39 mag in the 3.6, 4.5, 5.8, and 8.0 $\mu$m bands, respectively, and the appropriate aperture corrections were made using the values 
described in the IRAC Data Handbook\footnote{http://irsa.ipac.caltech.edu/data/SPITZER/docs/irac}.
To estimate the contamination due to non YSO candidates, we analyzed
a control field (Program ID 30734, AOR key 18902784: Galactic structure, PI: Steven Majewski) located approximately 1.1 degrees away from the target field with no obvious signs of star formation, using the same procedure as described above. For both the fields
 the sources which have photometric 
uncertainties less than 0.2 mag are considered as good detections. 
This should be considered as conservative lower limit as the calibration uncertainty
associated with the IRAC bands is 2-3\% and in addition, there are 
1-2\% uncertainty in the aperture corrections.
\begin{table*}
\centering
\scriptsize
\caption{Photometric data at IRAC bands. A sample of the table is given here. The complete table is available in electronic form.}
\begin{tabular}{cccccc}
\hline\hline
\multicolumn{1}{c} {RA (deg)} & \multicolumn{1}{c}  {DEC (deg)} &\multicolumn{1}{c}{[3.6]} &\multicolumn{1}{c}{[4.5]} & \multicolumn{1}{c}{[5.8]}& \multicolumn{1}{c}{[8.0]} \\
\multicolumn{1}{c} {J2000}  & \multicolumn{1}{c} {J2000} & \multicolumn{1}{c} {mag} & \multicolumn{1}{c} {mag} & \multicolumn{1}{c} {mag} & \multicolumn{1}{c} {mag} \\
\hline
109.124010 & -9.495950 & 11.942 $\pm$   0.009 & 11.927  $\pm$   0.012 & 11.942  $\pm$   0.023 & 11.949  $\pm$   0.076 \\
109.167320 & -9.489430 & 11.075 $\pm$   0.006 & 11.054  $\pm$   0.009 & 11.135  $\pm$   0.015 & 11.060  $\pm$   0.045 \\
109.142310 & -9.485300 & 9.981  $\pm$   0.004 & 10.008  $\pm$   0.005 & 9.995   $\pm$   0.009 & 9.875   $\pm$   0.040 \\
\hline
\end{tabular}
 \end{table*}
\section{Results}
\begin{figure}
\resizebox{8.5cm}{8.0cm}{\includegraphics{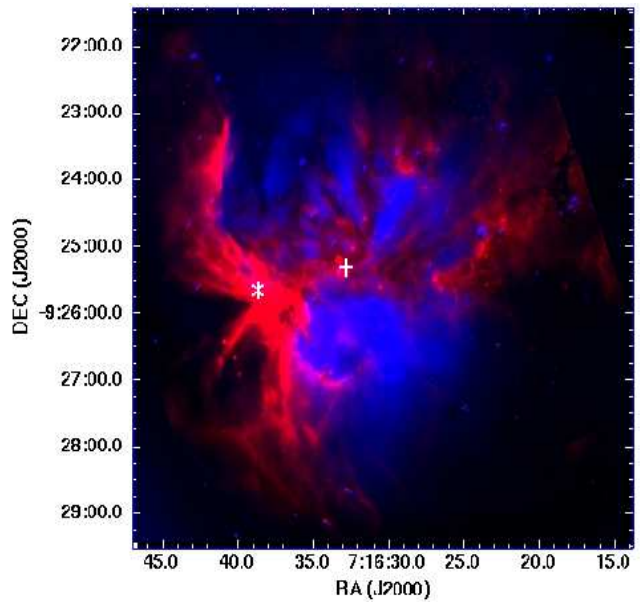}}
\caption{       {\it Spitzer}-IRAC 8 $\mu$m image
(red) superimposed on H$\alpha$ line image
(blue) of Sh2-294 region (for colour plot see online electronic version).  The image has a field of view $\sim$8\arcmin $\times$ 8\arcmin.
The plus and asterisk symbols show the positions of the
exciting source of Sh2-294 and of the massive B1.5 source  at its border, respectively.
 }
\label{NGC6823_field.ps}
\end{figure}
\subsection{Morphology of Sh2-294}
Figure 1 shows a colour composite image of H$\alpha$ emission (taken from Paper-I) superimposed with
the IRAC 8 $\mu$m  emission. The plus symbol in the image marks the position of the
exciting source of Sh2-294.  The position of the massive B1.5  source associated with ``region A'' is marked with an
asterisk symbol (see \S 1). The 8 $\mu$m IRAC band  contains emission bands at 7.7 and 8.6 $\mu$m commonly attributed to
PAH molecules. PAHs are believed to be destroyed in the ionized gas,  but
thought to be excited in the PDR at the interface of \hii region and molecular cloud by the absorption 
of far-UV photons leaking from the \hii region. Therefore, PAHs are good tracers of the
hot PDR that surrounds the \hii region. 

 Figure 1 also reveals the distribution of 8 $\mu$m emission in a filamentary fashion with its long axis aligned 
in east-west direction that bisects the \hii region and the ionized gas appears
to be streaming away orthogonal to the long axis. In both sides of the filament
axis, the ionized gas shows decrease in intensity possibly due
to the density  gradient in the original medium. The ionized gas is bounded
more in the eastern direction as compared to in western direction, however, the overall morphology of the nebula looks bipolar 
in nature. The bipolar morphology have been noticed in NIR to MIR bands 
in few \hii regions and/or bubbles (see e.g., Churchwell et al. 2006; Saito et al. 2009). Now long wavelength observation ($>$ 100 $\mu$m) with {\it Herschel} has added new insight to this morphology. For example, the recent  finding by Deharveng et al. (2012) with {\it Herschel} observations, the \hii region ``Sh2-201", illustrates 
a clear bipolar morphological nature, with two bipolar lobes perpendicular to a dense neutral filament extending east-west. The two lobes  seem to be bounded by cold neutral 
material of low  column density ($\leq$ 10$^{21}$ cm$^{-2}$), whereas the column density is high ($>$ 10$^{22}$ cm$^{-2}$) on each side at the waist of the Sh2-201 along the parental filament (Deharveng et al. 2012).
Bipolar \hii regions are  believed to be formed in a filamentary cloud. 
Due to the absence of sensitive long wavelength observations, it is hard to imagine the shape and structure of the original cloud, on a large scale in which Sh2-294 has formed. However, we shall discuss the importance 
\hii region evolution in a filamentary cloud as well as in a hierarchical clumpy cloud on the perspective of future star formation processes in Section 5.1.
 
\subsection{Identification of Young Stellar Objects}
The circumstellar dust emission from the disk and infalling envelope  of YSOs  gradually disappears with time.
The sequence of evolutionary phases of a YSO: embedded in a circumstellar envelope and an accretion disk 
(Class I), a classical T Tauri star surrounded by an optically thick disk (Class II), 
and a weak-line T Tauri star with an optically thin disk (Class III), are generally classified 
with slopes of SEDs between 2 to 20 $\mu$m (Lada \& Wilking 1984; Lada 1987). 
 The IRAC [3.6]-[4.5]/[5.8]-[8.0] and [3.6]-[4.5]/[4.5]-[5.8] colour-colour (CC) diagrams are also used as   
tools to identify  different classes of sources (cf. Allen et al.  2004; Hartmann et al. 2005) 
and follow a similar classification. Restricting our analysis to the sources having photometric error less than 0.2 mag, 
{ we  detected 102 sources  in  3.6, 4.5 and 5.8 $\mu$m  bands and 31 sources in all
the IRAC bands} in a common area of $\sim$6\arcmin.8 $\times$ 6\arcmin.8.
These catalogues are made after removing three sources that appear to be extended and/or  multiple sources in 
the high resolution H$_2$ (2.12 $\mu$m)  and  3.6 $\mu$m IRAC images. 
The final catalogue of the identified YSOs in the present analysis is given in Table 1. A sample of the
Table 1 is given here, whereas the complete table
is available in electronic from as part of the online material.
The H$_2$  mosaic image was constructed using the observations made with ISAAC camera of VLT. 
The data were retrieved from the ESO archive\footnote{http://archive.eso.org/eso/eso-archive-main.html}.
Further details about the observations are described in Yun et al. (2008).

The IRAC [3.6]-[4.5]/[5.8]-[8.0] CC diagram is shown in the left panel of Figure  2. Following the classification scheme by Allen et al. (2004) and Hartmann et al. (2005), the regions in the CC diagram occupied by Class II, Class I/II and Class 0/I objects are indicated. The sources with the colour of stellar photosphere are centered  around (0, 0); they include foreground, MS as well as Class III objects. The objects with 0.0 $\leq$ $[3.6]-[4.5]$ $\leq$ 0.8 and 
 0.4 $\leq$ [5.8]-[8.0] $\leq$ 1.1 are classified as Class II objects. The boundary for Class II sources 
is marked with a box and  are represented by  squares.
The dotted lines with [3.6]-[4.5] $\geq$ 0.7 and [5.8]-[8.0] $\geq$ 0.7 discriminate Class 0/I (triangles)
sources from Class II sources (Hartmann et al. 2005). The zone with  [5.8]-[8.0] $\ge$ 1.1 and [3.6]-[4.5] 
between 0.0 and 0.7 can be occupied by Class 0/I/II sources. Since Class 0 sources are generally not visible 
at wavelengths shorter than 10 $\mu$m, thus these objects are likely to be Class I/II sources and are shown with hexagon symbols.
On the basis of the above criteria, the [3.6]-[4.5]/[5.8]-[8.0] CC
diagram yields 20 YSOs. The search of YSOs on the basis of IRAC [3.6]-[4.5]/[5.8]-[8.0] 
CC diagram might have been affected by the lower sensitivity of the IRAC  8.0 $\mu$m band 
to detect faint sources and also by the bright background PAH emission.  
Therefore, in order to detect additional YSOs, we use IRAC [3.6]-[4.5]/[4.5]-[5.8] CC diagram shown in the right 
panel of Figure  2. 
Hartmann et al. (2005) show that 
the division between Class II and Class III T Tauri stars occurs at [3.6]-[4.5] and [4.5]-[5.8] $\sim$0.2 
with Class III stars are generally situated at [3.6]-[4.5] $<$ 0.2 and [4.5]-[5.8] $<$ 0.2, whereas  most of  the 
Class I/0 protostars are located at [3.6]-[4.5] $\ge$ 0.7 
and [4.5]-[5.8] $\ge$ 0.7. The region with [4.5]-[5.8] $\ge$ 0.7 and [3.6]-[4.5]  between 0.2 to 0.7  is
less defined and the overlap between Class I and Class II sources can be seen in well studied star-forming-regions 
(e.g., Muench et al. 2007; Koenig et al. 2008). On the basis of the above studies, we classify the sources 
that fall within or to the close vicinity of the boundaries marked
in [3.6]-[4.5]/[4.5]-[5.8] CC diagram. They are marked as Class II (squares), Class I (triangles) and 
Class I/II (hexagon) in Figure  2. It is to be noted that all the sources that 
show the characteristic of YSOs in IRAC [3.6]-[4.5]/[5.8]-[8.0] CC diagram are 
also identified as YSOs in [3.6]-[4.5]/[4.5]-[5.8] CC diagram.
However, it is possible that some of the sources identified as Class II  may be
reddened field stars/Class III sources.

The sample  of YSOs identified  on the basis  of MIR-CC diagram may be contaminated by
background dusty active galactic nuclei (AGN) and asymptotic giant branch (AGB) stars as 
they  have  colours similar to that of YSOs. In nearby SFRs, Fazio et al. (2004b) found that 
50\% of the sources with [3.6] mag $>$ 14.5 are extragalactic in nature.  
In our sample, we have only 3 common 
sources in the first three bands of IRAC, that have [3.6] mag $>$ 14.5. We 
visually inspected these sources in  VLT H$_2$ and {\it Spitzer} 3.6 $\mu$m images, and 
in both the images they appear as a point-like stellar object. 
We  used a  statistical approach to estimate the contamination  in the sample of the identified YSOs by comparing the MIR-CC  diagram of a control field (not shown) located  at $\alpha_{2000} = 07^{h}14^{m}49^{s}$,
$\delta_{2000} = -10^{\circ}27^{\prime}39^{\prime\prime}$ having an almost equal area to that  of the
Sh2-294 \hii region. Using the same diagnostic approach as described above, we find only 3 objects lying 
in the Class II and Class I zone of the MIR CC plot. This comparison suggests that the expected number contamination to the YSO population of Sh2-294 region is 
likely to be less than $10\%$.
\begin{figure*}
\centering
\resizebox{8.5cm}{8.5cm}{\includegraphics{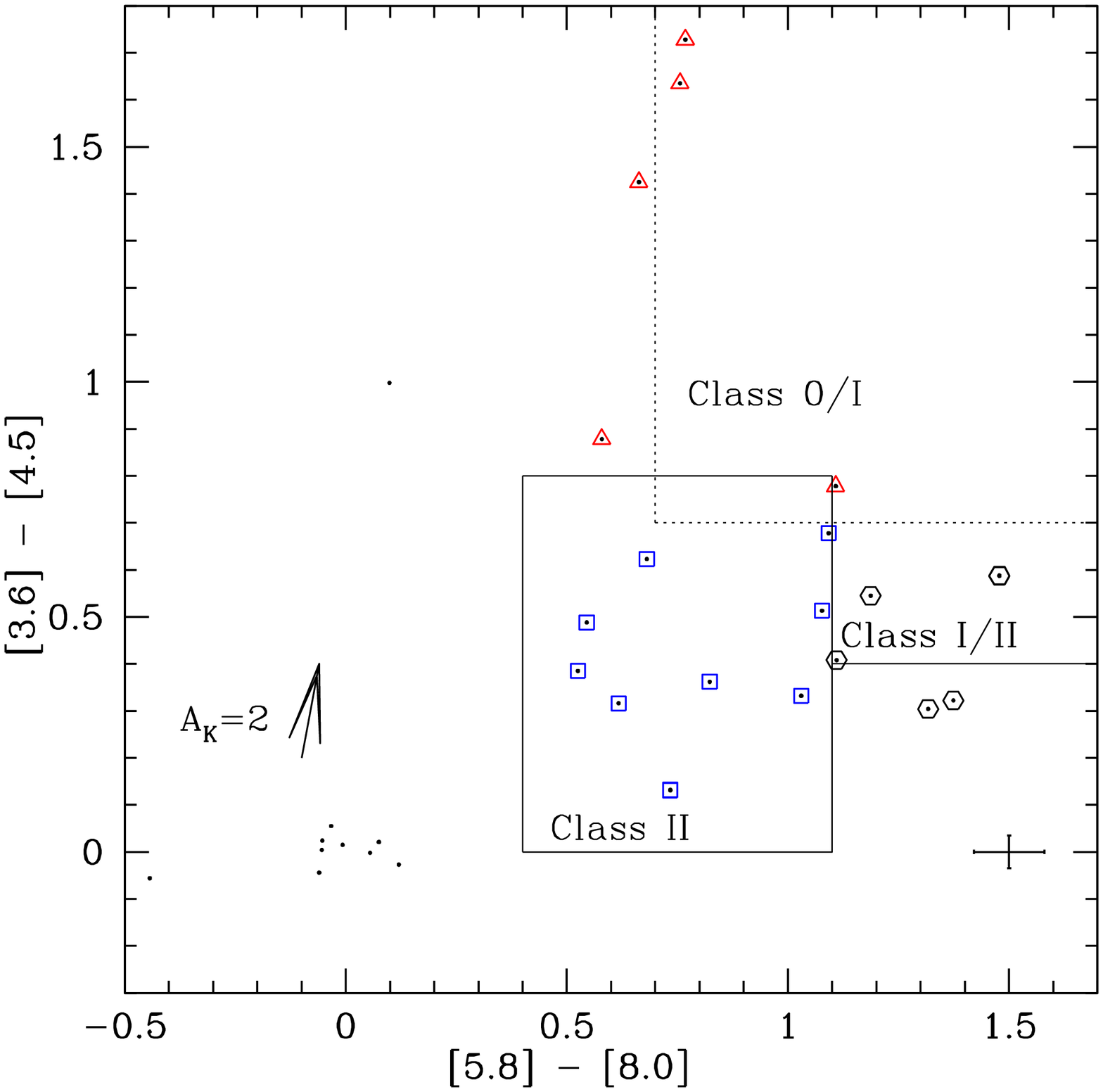}}
\resizebox{8.5cm}{8.5cm}{\includegraphics{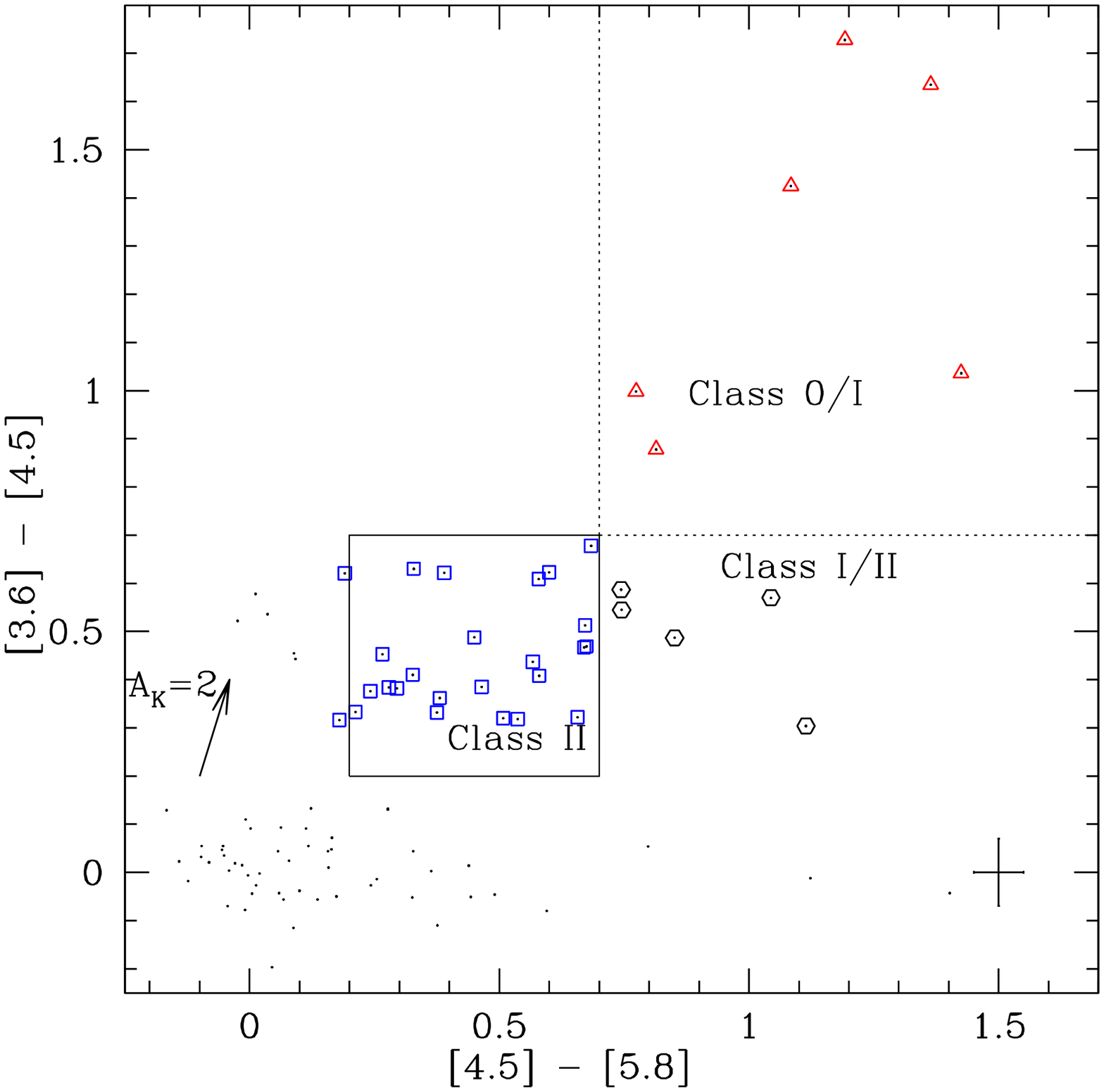}}
\caption{Left: IRAC [3.6]-[4.5]/[5.8]-[8.0] CC diagram with boxes representing the
boundaries of different class of sources. Right: IRAC [4.5]-[5.8]/[3.6]-[4.5] CC diagram
with boxes representing the boundaries of different class of sources (see text for details).
A reddening vector of A$_K$ = 2 mag, using extinction law of Flaherty et al. (2007) and mean error bars of the colours
are shown in both the IRAC CC diagrams. }
\label{NGC6823_field.ps}
\end{figure*}
\begin{figure}
\centering
\resizebox{8.5cm}{8.5cm}{\includegraphics{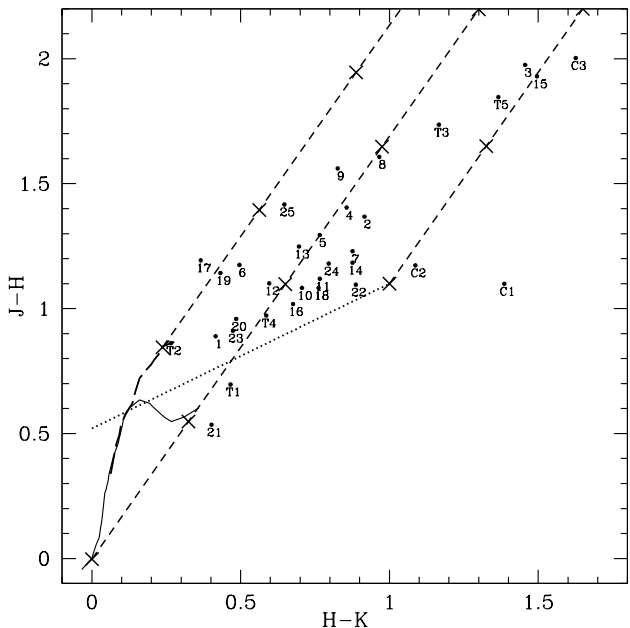}}
\caption{ NIR CC diagram for all the YSOs that have $JHK$ counterparts. The
thin solid line and
thick dashed curves represent the unreddened MS and giant
branches (Bessell $\&$ Brett 1988), respectively. The dotted line indicates the locus
of intrinsic CTTSs (Meyer et al. 1997). The curves and the colours are converted to the California Institute of Technology (CIT) system.
The parallel dashed lines are the reddening vectors drawn from the tip of the giant branch
(``upper reddening line''), from the base of the MS branch (``middle reddening line'') and
from the base of the tip of the intrinsic CTTS line (``lower reddening line''), with
crosses at every 5 mag of visual extinction.
The numbers 1-25, represent IRAC Class II sources, whereas   T1-T5 and C1-C3 denote to the
Class I/II and Class I sources, respectively. The magnitudes for C1 and C2 in $J$ and $H$ bands are the upper limit,
therefore their locations in the CC diagram are not the real. }
\label{NGC6823_field.ps}
\end{figure}
Figure 3 shows  $J-H$/$H-K$ CC diagram for the identified YSOs. The NIR counterparts were identified using a 
matching radius of $1\farcs22$.
In Figure  3, the IRAC classified Class II, Class I/II and Class I YSOs are labeled with the 
numbers 1-25, T1-T5 and C1-C3, respectively.  
Majority of the identified YSOs (barring T1 and 21) lie above the locus of the Classical T Tauri stars (CTTS) locus  given by Meyer et al. (1997), indicating that these could be the probable CTTSs. A 
comparison of the positions of these YSOs in the NIR CC diagram  (cf. Figure 3), with 
the 
CC diagram of the control field  (see Figure  9(b) of  Samal et al. 2007), 
again suggests that most of the sources identified using IRAC  CC 
diagrams are most likley YSO candidates. However, the positions of sources 17, 19, 21 and T2  in the
NIR CC diagram is not compatible with Class II sources.
The sources 17, 19 and T1 fall in the close vicinity of the leftmost reddening vector that starts from 
the tip of unreddened giant locus, whereas the position of the source 21  falls far below the T-Tauri locus.
The positions of these sources in the NIR CC diagram also mimic with the positions of the field star distribution of the control field.
Moreover, the spatial positions of three sources (17, 19 \& 21) show  that they are away from 
the cluster center (see \S 3.4).
 
From  IRAC photometry, we find a
total of 36 likely YSO candidates, indicating that star formation
 is still active in the region. 
It is not possible to distinguish Class III YSOs
from the MS and/or field stars with the existing photometric data alone.  Spectroscopic data and/or X-ray observations 
are needed to confirm the YSO nature of the Class III sources. The  X-ray emission in case of  
Class III stars is elevated by a factor of 10$^{3-4}$ compared to the MS stars (Walter \& Barry 1991). 
In the present analysis, we restrict ourself with these more reliable sample of Class I and Class II YSOs.
In the following sections, we have used
the spatial distribution of the identified YSOs and their characteristics to study the star 
formation scenario in Sh2-294 \hii complex.

\subsection{H$_2$ structures}
\begin{figure*}
\centering
\resizebox{10.0cm}{10.0cm}{\includegraphics{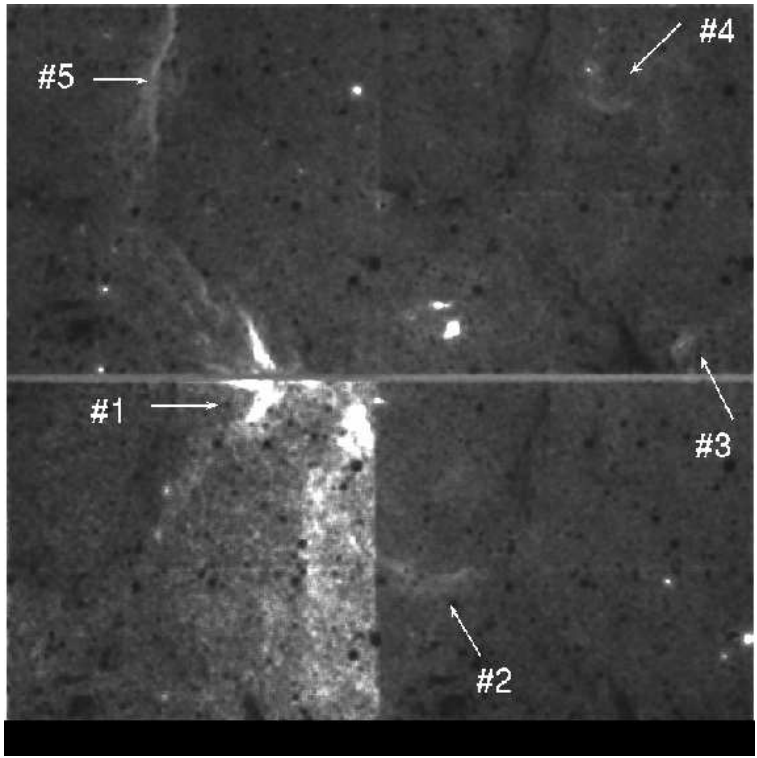}}
\caption {Median smoothed with the stars subtracted H$_2$ image at 2.12 $\mu$m.
The positions of the enhanced emissions are
marked as numbers. The point-like sources seen in the above smoothed image are the
product of the residuals that left in case of bright saturated stars.
{ The image is centered on $\alpha_{2000} = 17^{h}16^{m}32^{s}$,
$\delta_{2000} = -09^{\circ}25^{\prime}32^{\prime\prime}$ and has a field of view $\sim$5$^{\prime}$.2 $\times$ 5$^{\prime}$.2. North is up and east is to the left.}}
\label{NGC6823_field.ps}
\end{figure*}
\begin{figure*}
\center
\resizebox{16cm}{15cm}{\includegraphics{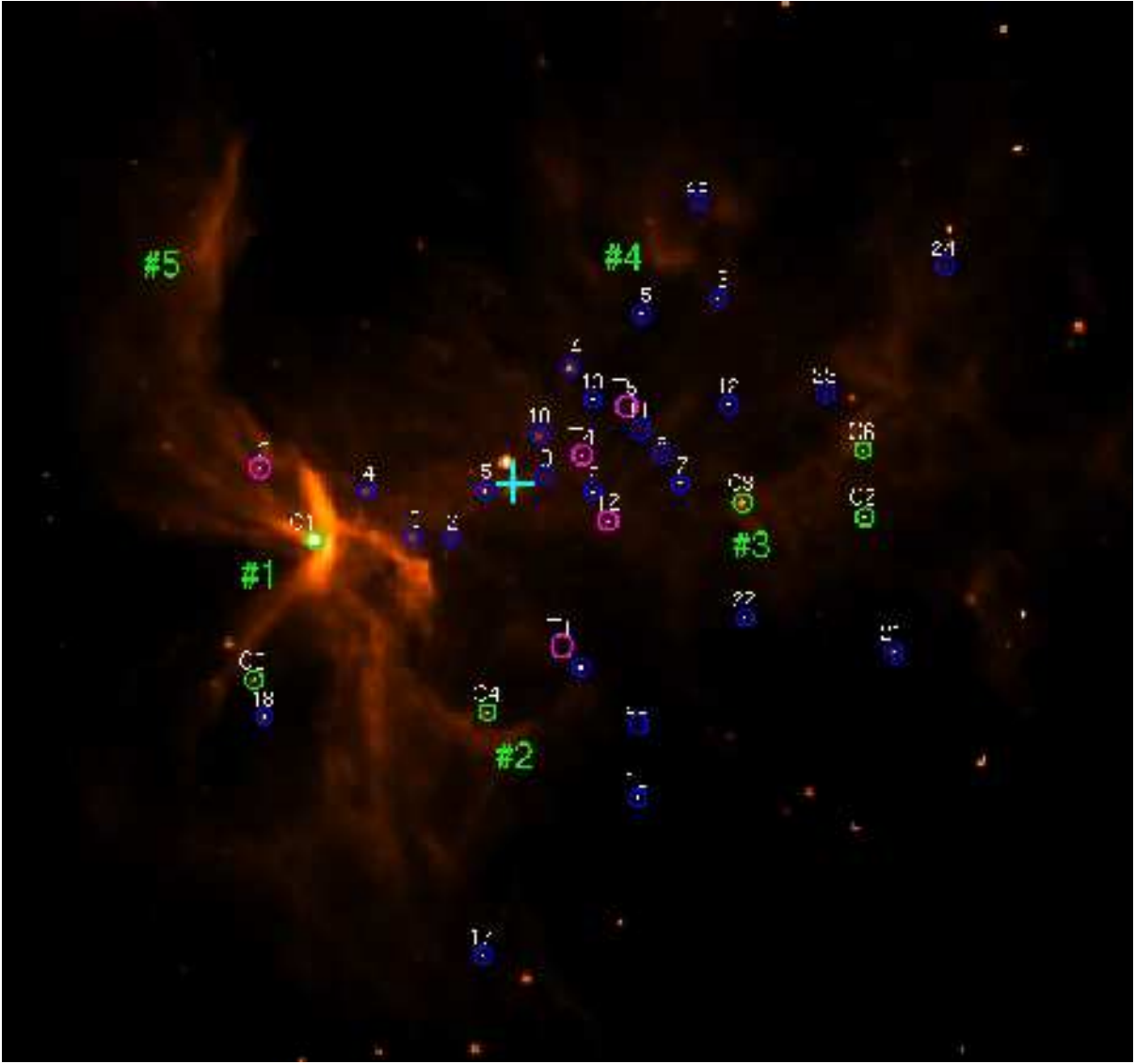}}
\caption { Spatial distributions of Class II (blue circles), Class I (green circles) and Class I/II (magenta circles)
overlaid on the 5.8 $\mu$m image (for colour plot see online electronic version).{ The image has a field of view $\sim$7$^{\prime}$.2 $\times$ 7$^{\prime}$.2 and is centered on $\alpha_{2000} = 17^{h}16^{m}32^{s}$, $\delta_{2000} = -09^{\circ}25^{\prime}32^{\prime\prime}$.}
 The YSOs are labeled in the same manner as in Figure  3. The plus symbol shows the position of the ionizing source. The enhanced H$_{2}$ structures are  leveled as in Figure 4. North is up and east is to the left.}
\label{NGC6823_field.ps}
\end{figure*}
The morphology of Sh2-294 region (see Figure 1) suggests that as the ionized gas streams away from the 
center, it possibly encounters an inhomogeneous  medium,  illustrating patchy, clumpy and knotty 
structures, which can be seen at {\it Spitzer} 8 $\mu$m emission. 
In order to correlate the structures seen in 8 $\mu$m, the positions of the detected YSOs and the H$_2$ emissions, we used the deep image at 2.12 $\mu$m, 1-0 
S(1) line of H$_2$, obtained with the VLT.  The present H$_2$ image is considerably more sensitive and 
has higher angular resolution than that presented in  Paper-I.
Figure  4 displays the H$_2$ image at  2.12 $\mu$m, which reveals
strong emission along the eastern border of the \hii region, as well as notable features at various locations of the nebula, 
which are numbered as $\#$1 to $\#$5. 
The diffuse H$_2$  emission can be
excited  by  Lyman  and  Werner  UV  photons  (e.g.,  Chrysostomou  et
al. 1992) from the massive  stars and thus can trace PDR or can be due to collisional excitation 
in the presence of shocks arising from outflows from nearby YSOs.  For   
dense  gas  ($\large{n}_{H}$ $\sim$ 10$^5$ cm$^{-3}$), H$_2$ emission alone cannot be  used to identify  PDR 
as UV excitation levels can be collisionally redistributed (see Allen et al. 1999), hence  another PDR  
tracer like PAH emission is  necessary to  trace PDR  around a nebula (Giard et al. 1994). Though, the 
spectroscopic observations of H$_2$ lines and their ratios would rather provide the exact nature of the excitation (see Luhman et al. 1998), however, close resemblance
 of diffuse H$_2$  emission  with the 8 $\mu$m emission (see Figure 1), suggests the H$_2$  emissions at 
the periphery of Sh2-294 are more likely caused by UV fluorescence. The identified H$_2$ structures are 
of particular interest as they show  rim or  arc-like morphology, with either  the rim faces
towards the ionizing source at the center or the curvature of arc (e.g., $\#$2) appears to be 
created  by the UV photons of ionizing star.
The morphology suggests that the weak H$_2$ and PAH emissions associated with the
Sh2-294 region possibly correspond to UV excited positions of the dense condensations or fragments of the original 
molecular cloud in which Sh2-294
was born.  However,  high resolution and high sensitive molecular line observations are needed to 
reveal the truth.  
Here, we presume that these  dusty structures are associated with the Sh2-294 SFR
because the rims/carved arcs are facing towards the ionizing star and exhibit  externally
heated structures, as expected in case of motionless cloud exposed to strong
UV radiation from  nearby massive star(s). 
\subsection{Spatial distribution of YSOs}
\begin{figure*}
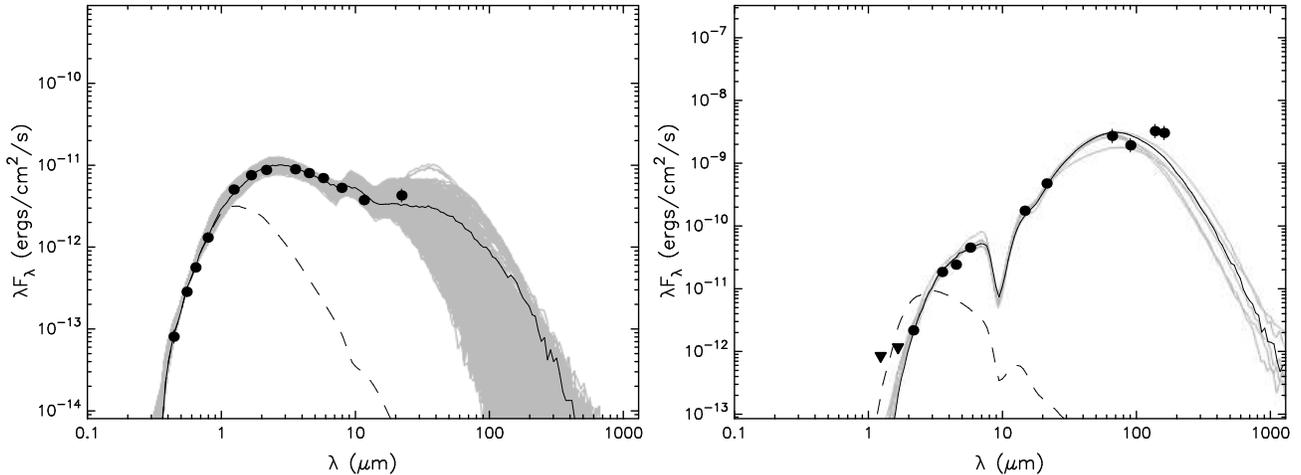

\centering{
\includegraphics[width=8.5cm] {Fig7a.eps}
\includegraphics[width=8.5cm] {Fig7b.eps}}
\caption{The SEDs for a Class II (Left) and  Class I (Right) source. (Left): The SEDs for the
 source 14. The black line shows
the best fit, and the grey lines show subsequent good fits with $\chi^2 - \chi^2_{\rm min} \leq 2N_{\rm data}$. The
dashed line shows the stellar photosphere corresponding to the
central source of the best fitting model. The filled circles denote the
input flux values.
(Right): The SED for the source C1. The lines are of same meaning as in Left figure. The filled circles denote the input flux values. The triangles are upper
limits for the flux values at $J$ and $H$ bands, taken from 2MASS.
The flux values at 14.65 and 21.30 $\mu$m are from $MSX$ point source catalog
(Egan et al. 2003) for the source G224.1880+01.2407 and the fluxes at 65, 90, 140 and 160 $\mu$m are from AKARI Far-Infrared Surveyor catlog
(Yamamura et al. 2009) for the source 0716378-092559.}
\label{FigVibStab}
\end{figure*}
We identified 36 YSOs (6 Class I, 25 Class II, and 5 Class InII) 
using IRAC CC diagrams; the spatial distribution of these YSOs on 
5.8 $\mu$m image is shown in Figure  5. The YSOs are labeled in the same manner as in Figure  3.
Figure  5 
indicates  that the YSO candidates are preferentially located in/around regions with 
diffuse 5.8 $\mu$m  emission. 
It is found that all the Class I sources are located at the outer regions,  whereas 
majority of  the Class II sources are found to be concentrated around the proximity of the ionizing source.  
If Class I sources are really the precursors to Class II sources, as expected 
from their less evolved circumstellar environments, they should have younger ages.
Hence, the distribution of Class I sources indicates that the 
star formation seems to be continued near the outer parts of the \hii region. 
The YSOs  C1, C2, C3, C4, C6, 1 and 25 are of particular interest because of their association with the
H$_2$ structures (as marked in Figure  5) and PAH emissions. 

The IRAC sensitivity and bright diffuse PAH emissions at
5.8 and 8.0 $\mu$m band could be one of the issues  for the detection of lower number of YSOs. 
We carried out the quantitative evaluation of the completeness of the photometric data 
with respect to the brightness and the position by adding artificial stars to
both 5.8 and 8.0 $\mu$m images. The ratio of the number of stars recovered to those added 
in each magnitude interval gives the completeness factor as a function of magnitude. We 
found that the completeness in the central region is $\sim$ 80\% to the level of $\sim$14.0 mag and $\sim$12.0 mag at 5.8 $\mu$m and 8.0 $\mu$m, respectively.
However, in the regions of diffuse emissions (e.g., $\#1$), the completeness decreases to 50\% at magnitudes $\sim$ 14.0
and $\sim$12.0 at 5.8 $\mu$m and 8.0 $\mu$m, respectively. Only one
Class I YSO has been identified in the region $\#1$  on the basis of IRAC data.
The  lower detection of YSOs could be the result of decrease in sensitivity due to the bright PAHs as obvious from the completeness analysis. 

\subsection{Physical properties of YSOs}
To get deep insight into the nature of the YSOs 
identified in the present work, we constructed SEDs using the models and fitting tools of 
Robitaille et al. (2006, 2007). The models are 
computed using a Monte-Carlo based radiation transfer code (Whitney et al. 2003a,b), which uses several combinations of central star, accreting disk, in-falling envelope and bipolar cavity 
for a reasonably large parameter space.  
Interpreting SEDs using the radiative transfer 
code is subject to degeneracies and spatially resolved multiwavelength
observations can break the degeneracy.
To constrain the parameters of stellar photosphere and circumstellar 
environment,  we  fit the SEDs 
to only those sources  for which we have a minimum of 6 data points in the
wavelength range from 0.55 to 8 $\mu$m. We search the counterparts of the 
identified YSOs in the optical-NIR bands by Yun et al. 2008 and 
in Two Micron All Sky Survey (2MASS) point source catalogue (Curti et al. 2003), using a matching radius of $1\farcs22$. 
This yields 33 sources detected in 6 bands or more ranging from 
optical ($VRI$) to IRAC (3.6, 4.5, 5.8, 8.0 $\mu$m) bands, which include 21 sources having optical counterparts. 
Out of 33 sources which have minimum 6 data points, 25 of them have fluxes at four wavebands of 
Wide-field Infrared Survey Explorer (WISE; Wright et al. 2010) survey. The WISE bands have central 
wavelengths at 3.4, 4.6, 12, and 22 $\mu$m, with  spatial resolutions of 
$6\farcs1$, $6\farcs2$, $6\farcs5$, and $12\farcs0$, respectively. As stated in Robitaille (2008), the SED 
fitting tool deals with a single source and input fluxes from multiple sources in a beam can result in incorrect 
stellar age and mass. Similarly, any erroneous fluxes can seriously affect the chances of obtaining sensible SED, thus sensible physical parameters.
As there is strong clustering of point sources seen in IRAC 4.5 $\mu$m image (not shown), along with  
bright nebulosities found in WISE 12 $\mu$m and 22 $\mu$m image, we therefore limited the use of WISE fluxes by 
preferring IRAC 3.6  $\mu$m and 4.5 $\mu$m fluxes over WISE 3.4  $\mu$m and 4.6 $\mu$m, due to 
superior spatial resolution of IRAC bands. Similarly, we use only those fluxes at 12 $\mu$m and 22 $\mu$m
 as inputs which have good-quality flags. We also set an upper limit at 22 $\mu$m for those sources for which 
we do not have detection,  by assigning the minimum 22 $\mu$m flux that found in our  sample of YSO. 
The SED fitting tool fits each of the models to the data, allowing the distance and external 
foreground extinction as free parameters.
Since we do not have 
spectral type information for our identified YSOs, in order to derive their approximate extinction values, we traced back 
these sources along the reddening vector to the intrinsic late MS locus or its extension in the  NIR CC diagram (e.g., Tej et al. 2006). 
Considering the uncertainties that might have gone into the estimates,
we used the estimated value \av~$\pm$ 2.5 mag, as an input parameter for these
sources. For Class I sources, for which we do not have NIR data, we allow 
\av~up to 30 mag. We further set 10$\%$ to 30$\%$ error in the flux estimates 
due to possible uncertainties in the calibration and intrinsic object variability. 
Figure  6 shows an example of SEDs of the resulting models for the
Class II and Class I sources. We obtained  physical parameters for 
all the sources adopting 
the approach similar to Robitaille et al. (2007) by considering those models that satisfy 
$\chi^2 - \chi^2_{\rm min} \leq 2N_{\rm data}$, where $\chi^2_{\rm min}$ is the goodness-of-fit parameter for the
best-fit model and $N_{\rm data}$ is the number of input observational
data points.
\begin{table*}
\centering
\scriptsize
\caption{Inferred Physical Parameters from SED Fits to YSOs}
\begin{tabular}{cccccccccc}
\hline\hline
 ID & \multicolumn{1}{c}{RA (deg)} & \multicolumn{1}{c}{DEC (deg)} &\multicolumn{1}{c}{$M_{\ast}$} &\multicolumn{1}{c}{$T_{\ast}$ } &\multicolumn{1}{c}{$t_{\ast}$} & \multicolumn{1}{c}{$M_{\rm disk}$ }
& \multicolumn{1}{c}{$\dot{M}_{\rm disk}$ } & \multicolumn{1}{c}{A$_V$ } & ${\chi}^2$ \\

 &\multicolumn{1}{c}{(J2000)} & \multicolumn{1}{c}{(J2000)} &\multicolumn{1}{c}{($M_\odot$)}
          & \multicolumn{1}{c}{(10$^{3}$ K)}
          & \multicolumn{1}{c}{(10$^{6}$ yr)} & \multicolumn{1}{c}{( $M_\odot$)} & \multicolumn{1}{c}{(10$^{-8}$ $M_\odot$/yr)} &
          \multicolumn{1}{c}{mag}\\ \hline
\hline
1 & 109.130424 & -9.443050 & 1.99       $\pm$   0.36 & 5.83     $\pm$   1.23 & 4.85     $\pm$   2.64 & 0.009    $\pm$   0.012 & 3.97    $\pm$   3.66 & 3.82     $\pm$   0.59 & 2.82\\
2 & 109.145180 & -9.428912 & 2.19       $\pm$   0.37 & 8.68     $\pm$   1.64 & 6.99     $\pm$   2.42 & 0.006    $\pm$   0.013 & 3.44    $\pm$   3.42 & 8.13     $\pm$   0.53 & 0.89\\
3 & 109.149498 & -9.428577 & 3.00       $\pm$   0.47 & 11.03    $\pm$   1.80 & 5.10     $\pm$   2.41 & 0.003    $\pm$   0.009 & 1.12    $\pm$   1.61 & 13.59    $\pm$   1.15 & 8.05\\
4 & 109.154785 & -9.423507 & 2.53       $\pm$   0.53 & 8.01     $\pm$   2.13 & 4.45     $\pm$   1.33 & 0.004    $\pm$   0.007 & 4.47    $\pm$   4.54 & 8.72     $\pm$   0.50 & 0.98\\
5 & 109.141357 & -9.423445 & 2.87       $\pm$   0.65 & 9.75     $\pm$   3.12 & 4.99     $\pm$   2.43 & 0.005    $\pm$   0.016 & 2.30    $\pm$   2.77 & 7.61     $\pm$   0.73 & 1.05\\
6 & 109.128922 & -9.423709 & 2.24       $\pm$   0.35 & 9.01     $\pm$   1.57 & 7.66     $\pm$   1.62 & 0.001    $\pm$   0.003 & 0.84    $\pm$   0.86 & 7.35     $\pm$   0.67 & 4.07\\
7 & 109.118996 & -9.422717 & 2.25       $\pm$   0.34 & 9.14     $\pm$   1.52 & 7.55     $\pm$   1.34 & 0.001    $\pm$   0.004 & 0.75    $\pm$   0.81 & 7.56     $\pm$   0.52 & 20.35\\
8 & 109.134392 & -9.421776 & 2.44       $\pm$   0.55 & 7.65     $\pm$   2.72 & 4.47     $\pm$   2.79 & 0.007    $\pm$   0.016 & 3.36    $\pm$   4.55 & 10.90    $\pm$   0.90 & 2.28\\
9 & 109.121185 & -9.419425 & 3.50       $\pm$   0.83 & 7.49     $\pm$   4.17 & 1.56     $\pm$   0.98 & 0.011    $\pm$   0.018 & 10.83   $\pm$   10.68 & 10.03   $\pm$   1.50 & 4.85\\
10 & 109.135048 & -9.417233 & 4.85      $\pm$   0.31 & 16.10    $\pm$   1.03 & 4.93     $\pm$   0.98 & 0.000    $\pm$   0.001 & 0.04    $\pm$   0.04 & 6.64     $\pm$   0.30 & 7.30\\
11 & 109.123581 & -9.416539 & 2.31      $\pm$   0.48 & 5.69     $\pm$   1.03 & 3.23     $\pm$   2.24 & 0.010    $\pm$   0.019 & 2.80    $\pm$   2.65 & 6.38     $\pm$   0.59 & 1.45\\
12 & 109.113495 & -9.413931 & 1.98      $\pm$   0.33 & 5.15     $\pm$   0.84 & 3.17     $\pm$   2.30 & 0.006    $\pm$   0.015 & 1.35    $\pm$   4.10 & 5.45     $\pm$   0.45 & 1.92\\
13 & 109.128975 & -9.413376 & 2.28      $\pm$   0.27 & 9.48     $\pm$   1.19 & 8.09     $\pm$   1.15 & 0.002    $\pm$   0.005 & 0.16    $\pm$   0.16 & 7.63     $\pm$   0.46 & 8.45\\
14 & 109.131577 & -9.409706 & 3.09      $\pm$   0.33 & 11.14    $\pm$   1.26 & 3.71     $\pm$   1.64 & 0.013    $\pm$   0.019 & 1.43    $\pm$   1.42 & 6.33     $\pm$   0.16 & 4.45\\
15 & 109.123474 & -9.403764 & 4.64      $\pm$   1.01 & 4.65     $\pm$   0.27 & 0.17     $\pm$   0.13 & 0.007    $\pm$   0.017 & 3.88    $\pm$   3.88 & 13.49    $\pm$   0.39 & 17.58\\
16 & 109.114693 & -9.401979 & 3.16      $\pm$   0.63 & 11.39    $\pm$   2.68 & 5.24     $\pm$   1.99 & 0.002    $\pm$   0.010 & 2.85    $\pm$   3.59 & 4.85     $\pm$   0.58 & 2.38\\
17 & 109.141520 & -9.475020 & 0.60      $\pm$   0.79 & 3.46     $\pm$   0.59 & 0.10     $\pm$   0.13 & 0.007    $\pm$   0.008 & 28.69   $\pm$   28.04 & 5.40    $\pm$   0.27 & 8.85\\
18 & 109.166450 & -9.448670 & 1.89      $\pm$   0.53 & 6.59     $\pm$   2.74 & 3.32     $\pm$   3.21 & 0.036    $\pm$   0.028 & 52.65   $\pm$   39.97 & 4.95    $\pm$   0.96 & 4.35\\
19 & 109.123810 & -9.457560 & 2.57      $\pm$   0.74 & 8.61     $\pm$   3.76 & 4.66     $\pm$   1.20 & 0.000    $\pm$   0.000 & 0.06    $\pm$   0.05 & 6.22     $\pm$   1.13 & 1.95\\
20 & 109.123770 & -9.449440 & 1.88      $\pm$   0.46 & 5.20     $\pm$   0.40 & 4.82     $\pm$   2.52 & 0.007    $\pm$   0.011 & 1.15    $\pm$   1.05 & 4.34     $\pm$   0.32 & 3.15\\
21 & 109.094720 & -9.441330 & 4.79      $\pm$   0.10 & 16.04    $\pm$   0.21 & 4.07     $\pm$   0.28 & 0.000    $\pm$   0.000 & 0.00    $\pm$   0.00 & 3.10     $\pm$   0.34 & 11.15\\
22 & 109.111640 & -9.437610 & 3.57      $\pm$   0.38 & 12.30    $\pm$   3.08 & 5.73     $\pm$   2.30 & 0.000    $\pm$   0.002 & 0.07    $\pm$   0.07 & 6.75     $\pm$   1.31 & 25.51\\
23 & 109.102440 & -9.412630 & 3.60      $\pm$   0.30 & 12.38    $\pm$   3.06 & 5.91     $\pm$   2.26 & 0.000    $\pm$   0.000 & 0.03    $\pm$   0.03 & 6.76     $\pm$   1.32 & 6.45\\
24 & 109.088880 & -9.398560 & 1.22      $\pm$   0.54 & 4.18     $\pm$   0.58 & 0.07     $\pm$   0.28 & 0.038    $\pm$   0.034 & 37.83   $\pm$   37.80 & 6.20    $\pm$   0.76 & 10.51\\
25 & 109.116930 & -9.391600 & 3.62      $\pm$   0.29 & 12.97    $\pm$   1.09 & 4.29     $\pm$   1.94 & 0.000    $\pm$   0.000 & 0.00    $\pm$   0.00 & 5.81     $\pm$   0.84 & 8.67\\
T1 & 109.132507 & -9.440776 & 2.27      $\pm$   1.22 & 4.97     $\pm$   0.59 & 3.10     $\pm$   2.38 & 0.014    $\pm$   0.015 & 7.91    $\pm$   7.90 & 5.33     $\pm$   0.62 & 13.29\\
T2 & 109.127251 & -9.426885 & 2.32      $\pm$   0.58 & 6.90     $\pm$   1.91 & 4.37     $\pm$   2.55 & 0.011    $\pm$   0.017 & 34.07   $\pm$   33.95 & 3.34    $\pm$   0.71 & 2.63\\
T3 & 109.166985 & -9.420988 & 5.87      $\pm$   0.96 & 5.10     $\pm$   0.65 & 0.16     $\pm$   0.12 & 0.052    $\pm$   0.086 & 149.58  $\pm$   149.45 & 11.40  $\pm$   1.01 & 6.23\\
T4 & 109.130173 & -9.419502 & 1.82      $\pm$   0.90 & 5.39     $\pm$   2.27 & 0.80     $\pm$   1.26 & 0.026    $\pm$   0.028 & 73.84   $\pm$   73.34 & 10.08   $\pm$   0.83 & 8.87\\
T5 & 109.125168 & -9.414126 & 0.76      $\pm$   0.42 & 4.00     $\pm$   0.43 & 0.29     $\pm$   1.09 & 0.004    $\pm$   0.009 & 4.28    $\pm$   4.27 & 3.54     $\pm$   0.35 & 7.74\\
C1 & 109.162000 & -9.428901 & 8.78      $\pm$   1.67 & 6.12     $\pm$   1.61 & 0.04     $\pm$   0.04 & 0.199    $\pm$   0.274 & 390.00  $\pm$   383.00 & 23.70  $\pm$   6.69 & 35.89\\
C2 & 109.098038 & -9.426320 & 2.81      $\pm$   0.78 & 4.51     $\pm$   0.19 & 0.17     $\pm$   0.09 & 0.008    $\pm$   0.018 & 6.94    $\pm$   6.94 & 6.16     $\pm$   1.35 & 10.15\\
C3 & 109.111923 & -9.424769 & 2.66      $\pm$   1.93 & 5.82     $\pm$   4.75 & 0.37     $\pm$   1.03 & 0.034    $\pm$   0.048 & 112.23  $\pm$   110.78 & 14.17  $\pm$   1.21 & 18.19\\
\hline
\end{tabular}
 \end{table*}
  The parameters are obtained from the 
weighted mean and standard
deviation of these  models,  weighted by
e$^{({{-\chi}^2}/2)}$ of each model and are tabulated in Table 2. These parameters are obtained from set of models that 
represent the overall distribution, therefore more likely suppress any extreme values which may arise due to few badly fitted
models. However, due to limited observational data points, some of the parameters from the model fits can be  narrowly constrained over others depending upon the
available fluxes. Since we are mainly interested in stellar parameters due to availability of shorter wavelength fluxes,  we therefore only quote few parameters from the 14 parameter space SED models.
Table 2 lists the star  mass ($M_\star$), temperature ($T_\star$), stellar age ($t_\star$), mass of the disk 
($M_{\rm disk}$), disk accretion rate ($\dot{M}_{\rm disk}$), foreground visual absorption ($A_V$) 
and the $\chi^2_{\rm min}$ of the best fit.
Table 2 reveals that the ages of the majority of Class II  YSO (excluding 17) candidates vary
between 1.5 to 7 $\times$ 10$^{6}$ yr with a median age of 4.5 $\pm$ 1.9 
$\times$ 10$^{6}$ yr, whereas the visual 
extinction varies from 3.2 to 13.5 mag, indicating presence of non-uniform extinction within the region. 
 Out of 25 Class II sources, 18 sources have fluxes at optical wavelengths, thus their age estimations expected to be better  constrained, which  is 5.1 $\pm$ 1.4 $\times$ 10$^{6}$ yr (median age). From Table 2, 
it appears that the YSOs identified in the present work are detected nearly down to $\sim$ 1.8 
 $\msun$ and probably a large number of low mass YSOs are embedded in the cloud.
We note that the stellar ages given in Table 2 are only 
approximate. Spectroscopic observations of the YSOs would be more accurate in the determination
of their stellar age.  In Robittalle et al. (2007) models, the stellar masses and ages are sampled using the
isochrones and evolutionary tracks (Bernasconi \& Maeder 1996; Siess et al. 2000) from  the stellar luminosity 
and temperature, derived from the SED models. Since the age determination for all the YSOs is being
done following the same set of evolutionary tracks and the same approach, we  therefore use  the relative 
age among the YSOs to constrain the star formation history of the complex. However, we are aware of the 
fact that age estimation depends upon the choice of isochrones. For example, Hillenbrand et al. (2008) suggested  use of 
different set of isochrones can lead to a systematic uncertainty at a level of 0.75 dex for sub-solar mass stars, though the
agreement is better for older PMS solar-mass stars.  Out of  six Class I sources identified in the present study,  only three (C1, C2 \& C3)  have sufficient data points to fit the models of Robitaille et al. (2007). The models predict that  all the  three sources are young ($\leq$ 0.4 $\times$ 10$^{5}$ yr) and have relatively massive disk in comparison to the  Class II sources of the region.  The age estimation of the accreting embedded protostars probably more uncertain, therefore, should considered as quantitative indicator of stellar youth. Though, 
we do not have sufficient data to address the evolutionary status of
other  Class I YSOs (C4, C5 \& C6), however, the infrared colours of these objects indicate that they must be IR excess stars with circumstellar disk.  Here, we assume the age of C4, C5 and C6 must be of
the order of few $\times$ 10$^{5}$ yr, which is generally attributed to this 
class of sources (Kenyon \& Hartmann 1995). As discussed earlier,  most of the Class I sources
 are distributed away from the central cluster.
The Class I sources represent a much younger population than the ionizing source ($\sim$4 $\times$ 10$^{6}$ yr)
and the associated Class II sources within the cluster region.  It is also apparent from  Table 2 that 
the disk accretion rate of the  Class II YSOs, except few outliers, is in
the range of 10$^{-7}$ $-$ 10$^{-9}$ \msun~yr$^{-1}$, which is less than that of 
Class I/II and Class I YSOs (10$^{-6}$ $-$ 10$^{-7}$ \msun~yr$^{-1}$). 
 It is to be noted that in the absence of far-infrared (FIR)  to millimeter data the above values, particularly 
the disk parameters, should be treated with a caution.
\subsection{Massive YSO}
Samal et al. (2007) on the basis of mid to far-infrared fluxes predicted a presence of a massive ZAMS
star in the region $\#1$. In the {\it Spitzer} image, we identified a luminous source (C1; [3.6]=12.1 mag)
located behind the rim like structure seen in $5.8$ $\mu$m image (see Figure 5). 
The star is faint in the K$_s$ band and do  not has stellar counterpart in the optical. The SED models of the source based on data from 2MASS (J, H, and K$_s$), {\it Spitzer} (3.6, 4.5, and 5.8 $\mu$m), MSX survey (14.65  and 21.30 $\mu$m), and AKARI (65, 90, 140 and 160 $\mu$m) are shown in Figure. 6 (right panel). %
The SED of this YSO candidate is better sampled at longer wavelengths, showing a very steep rise in the 1.22 $\mu$m to $\sim$ 140 $\mu$m range, indicating the presence of an envelope thus the object should be a Class I YSO. The parameters from the best fit models suggest that the source has an age
of $\sim$4 $\times$ 10$^{4}$ yr and mass $\sim$9 \msun~ with a total luminosity 
of $\sim$1.8 $\times$ 10$^{3}$  \lsun. The source is
embedded in a cloud having \av~$\sim$24 mag and is still
accreting with a high envelope accretion rate of $\sim$9.8 $\times$ 10$^{-4}$ \msun~yr$^{-1}$. Accreting 
protostars show signature of outflow in their very young age.  Extended 4.5 $\mu$m emission often used as 
a tracer of outflow activity from massive young stellar objects (MYSOs), as the 4.5 $\mu$m IRAC band contains both \h\/ (v=0-0, S(9,10,11)) lines and CO
(v=1-0) band heads, thus can be excited by shocks such as those expected from protostellar 
outflows (Cyganowski et al. 2008 and references therein). Outflow sources identified based on their enhanced extended 
 4.5 $\mu$m emission are known as ``Extended Green Objects (EGOs)" and are generally identified with color coding using IRAC  
3-colour composite images. In search of EGOs, we made a colour-composite (see Figure 7)
image using the IRAC first three bands around the MYSO ``C1". The source is embedded in a PDR
and PDRs are generally bright in IRAC 5.8 $\mu$m and 8.0 $\mu$m, thus adds difficulty in search for weak EGOs. 
In Figure 7, however, we did not find  any enhanced emission at 4.5 $\mu$m, indicating either the 
emission is absent or very weak.
It has been noticed that the YSOs responsible for outflows appear blue in IRAC [3.6]-[4.5]/[4.5]-[5.8] CC diagram and are generally 
located at specific position ([3.6]-[4.5] $\geq$ 1.1 mag and [4.5]-[5.8] $\leq$ 0.8 mag; Ybarra \& Lada 2009 and references therein) in this CC space; thus sometimes used with caution to identify the 
driving sources. A comparison shows that none of the  YSO candidates in our sample are located at the preferred position as mentioned above, in
IRAC [3.6]-[4.5]/[4.5]-[5.8] CC diagram (see Figure 2$b$), indicating the identified YSOs are probably crossed the outflow stage.
A star of  $\sim$9 \msun~takes $\sim$2 $\times$ 10$^{5}$ yr to reach ZAMS (Bernasconi \& Maeder 1996) and the 
evolutionary status (age $<$ 10$^{5}$ yr) of the MYSO  based on SED models indicate that the source is still the 
in PMS phase or about to reach ZAMS. In the PMS phase, the Lyman continuum emission 
from a massive star is expected to be much lower than 
its corresponding MS phase. In Paper-I, we found a radio continuum
peak in the proximity of structure  $\#1$; however, the emission is elongated 
along the bright arc seen in the 5.8 $\mu$m image. The {\it Spitzer} image reveals
that massive source C1 does not exactly coincide with the radio peak, it is rather situated 
$\sim$20\arcs~ away in the eastern direction. Though the resolution of the radio continuum image 
presented in Paper-I, does not allow to conclude about the origin of the radio emission,
however, we presume that its elongated structure and its offset from the source 
C1  possibly represent the ionized gas from the photo-evaporating layer of 
the externally ionized rim (e.g., Morgan et al. 2004). The high envelope 
accretion rate indicates that source C1 may become more massive with time and may drive an \hii region. 
\begin{figure}
\centering
\resizebox{8cm}{8cm}{\includegraphics{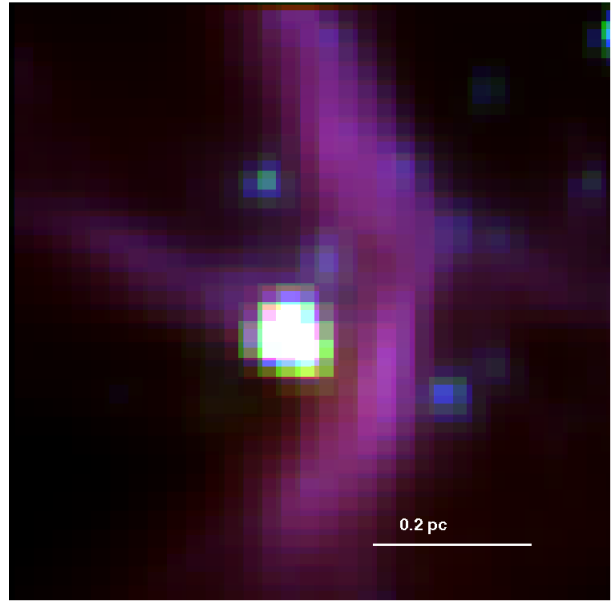}}
\caption{The {\it Spitzer}-IRAC colour-composite image (5.8 $\mu$m, red; 4.5 $\mu$m, green; and  3.6 $\mu$m, blue) around the massive source C1. North is up and east is to the left.
 }
\label{NGC6823_field.ps}
\end{figure}

\section{Discussion on star formation activity}
Star formation in a molecular cloud is more active during the first few million year of cloud 
lifetime. It is believed that majority of the stars in a molecular cloud form in clusters and  after few million year star formation, 
the gas in a molecular cloud dissipates and further star formation 
no longer takes place (Lada \& Lada 2003).
It is found that clusters with age greater than  $\sim$ 5 $\times$ 10$^{6}$ yr are seldom associated with molecular gas (Leisawitz et al 1989), thus it is more likely that after $\sim$ 5 $\times$ 10$^{6}$ yr no gas is left over to form new stars.
However, these type of molecular cloud complexes can still have star formation in PDRs 
at the interface of the \hii region and the molecular gas.
Many young clusters associated with \hii regions show age spread, which is 
partly due to the cluster evolution and partly due to different epochs of 
star formation within the region (e.g., Sharma et al. 2007; Pandey et al. 2008; 
Jose et al. 2008).
In Sh2-294 region, with the help of {\it Spitzer} observations, we detected sources with disk and envelope, 
evidence of youth of the region. The presence of extremely young 
Class I sources perhaps represents fresh star formation in the region. 
The SED models, within the uncertainties, indicate that the age of most of the Class II YSOs are of comparable to ionizing source and these sources are 
certainly older than than Class I YSOs.
All the Class I sources are distributed at the 
outskirts of the Sh2-294 region; most of them are associated with arc-shaped 
H$_{2}$ structures and PAH features.
 The distribution of young YSOs at the outskirts of \hii
regions/bubbles has been noticed in several cases (see e.g., Deharveng et al. 2005; Zavagno et al. 2006; Koeing et al. 2008; Watson et al. 2008; Deharveng et al. 2009; Chauhan et al. 2011), where it is believed that 
the majority of them are formed as a result of triggered
star formation. 
Although  kinematics of the Sh2-294 region is
not available,  the morphology, the association
of Class I YSOs with  H$_2$ structures and their younger ages with respect to 4 $\times$ 10$^{6}$ yr massive B0 MS star, give evidence in favor of  triggered star formation in the region.
On the basis of Hipparcos data, Madsen et al. (2002) calculated
the velocity dispersion of $\sim$1 km s$^{-1}$ for stars in young clusters 
and associations. Since the Class I sources are 
expected to be young ($\sim$10$^{5}$ yr), their positions should roughly indicate the place where they have born, whereas with a velocity dispersion of 
1 km s$^{-1}$, the Class II YSOs (median age $\sim$4.5 $\times$ 10$^{6}$ yr) 
might have drifted away $\sim$ 4 pc from their original position. 
Therefore, some of the Class II YSOs distributed away from 
the cluster center ($\alpha_{2000}=07^{h}16^{m}33^{s}$,
$\delta_{2000}=-09^{\circ}25^{\prime}35^{\prime\prime}$; see Samal et al. 2007) might have drifted due to their motion.
Figure 5 shows the spatial distribution of YSOs.
The Class I source C4 is associated with the structure $\#2$ and is
situated at the vertex  of a finger-like structure pointing
towards the ionization source. A similar kind of example can be seen  in the 
case of RCW 120 (Zavagno  et al. 2007; Deharveng  et al. 2009), where a YSO 
located at the vertex of a structure pointing towards the exciting star (Deharveng  et al. 2009).
Deharveng  et al. (2009) proposed that the structure results from the dynamical instability of 
the expanding ionization front, such as those simulated by Garcia-Segura \& Franco (1996).
The regions $\#1$ and  $\#3$ are associated with Class I sources.
The region $\#1$ is associated with a relatively  massive ($\sim$9 \msun) star and is more prominent in PAH emission in comparison to the region $\#3$, which is associated with a less massive ($\sim$2.7 \msun) star.
The  Class II source  25 seems to be projected near the H$_{2}$ 
structure $\#4$, but its parameters derived from the acceptable SED fitting models suggests 
that the source is possibly not embedded within the structure.

Though the spatial distribution and evolutionary status of YSOs suggest in favor of triggered star formation, however, 
with the present data  it is not possible to pinpoint the exact nature of triggering, but it is worth discussing
the two most likely scenarios  that might have happened in this region. In the the first 
case, we discuss the star formation as seen in remnants of pre-existing molecular material of
a parental clumpy cloud due to interaction of an expanding \hii 
region. In the second case, we discuss the  evolution of an \hii region in a filament and the resulting
star formation.

In Sh2-294 region, we have identified structures ($\#1$ and $\#3$) with YSO inside and they resembles
with the  numerical simulation of globules resulted due to the impact of UV photons from  the near-by massive OB stars 
on the pre-existing dense molecular material (Lefloch \& Lazareff 1994). The structure  $\#1$ shows clearly
a externally  heated rim, with a massive (9 \msun) source C1 inside the rim, whereas the region $\#3$ has a more globular morphology, 
with a less massive (2.7 $\msun$) source C3 inside. In both the cases, the structures show cometary morphology with their heads pointing toward the  illuminating B0 MS star. Both the structures are approximately at the same distance form the ionizing source. The different morphological structures under the same set of initial conditions and exposed to approximately the same amount of ionizing photons, may be due to the differences in size and mass of the pre-existing clumps. Several studies show the association of  YSOs with bright rims at the borders of \hii regions (Sugitani \& Ogura 1994; Ogura et al. 2002)
and their formation is more likely due to effect of strong external radiation from OB stars (Morgan et al. 2004;
Urquhart et al. 2006) . On the basis of cometary morphology, location, and age difference 
between the associated YSOs and the ionization source, one can anticipate that 
radiation driven implosion (RDI; Lefloch \& Lazareff 1994; Miao et al. 2006) could be
the process of star formation for regions $\#1$ and $\#3$.
To evaluate this hypothesis we quantitatively compare 
the time elapsed during the formation of the Class I YSO (C1)  in the region $\#1$ with that of the age of the B0 ionization source at center.
In case of RDI-induced star-formation, the  pre-existing molecular clumps are surrounded 
by high-pressure ionized gas due to the photoionizing UV photons 
and the pressure exerted on the surface of the molecular clump 
leads to formation of a cometary globule. At an appropriate time, the 
high pressure will drive a shock front into the clump, leading to the 
formation of new  stars (for details, see  Lefloch \& Lazareff 1994). We estimated the time needed for the ionization front (IF) to 
travel to the present position of the rim of the structure  $\#1$ situated at a projected distance 
of 1.7 pc as $\sim$1.5 $\times$ 10$^{5}$ yr, assuming that the IF expands at the sound 
speed of 11.4 km s$^{-1}$. 
The characteristic timescale for producing cometary 
morphologies of various shapes and inducing gravitational collapse varies from 0.1 to $\sim$1 $\times$ 10$^{6}$ yr 
(Lefloch \& Lazareff 1994; Miao et al. 2009), depending upon initial conditions.Hence, 
we presume an elapsed time to initiate star formation inside the region $\#1$ could be $\leq$ 1 $\times$ 10$^{6}$ yr.
The age of the Class I source inside the rim should be of the order 
of 10$^{5}$ yr.
Taken together,  the summed time scale 
($\sim$2.6 $\times$ 10$^{6}$ yr) is less than the age 
($\sim$4.0 $\times$ 10$^{6}$ yr) of the ionizing source that powers 
the \hii region.
We also  evaluated the shock crossing time to the globule to see whether the star formation in regions 
$\#1$ and $\#3$ started due to the propagation of ionization-shock front or whether it  had already taken place  prior to the
ionization front arrival. Assuming a typical shock velocity of 1-2 km s$^{-1}$, as found in the case of Bright Rimmed Clouds  (BRCs; see e.g., Thompson et al. 2004; White et al. 1999) for
the neutral gas associated with $\#1$, the shock
travel time to the source C1 projected at a distance $\sim$ 0.31 pc from the 
photo-ionization surface layer ranges from  3.0-1.5 $\times$ 10$^{5}$ yrs. 
This time scale is comparable to the age of the source C1. These simple 
approximate estimations suggest that the formation of the source C1 and similarly C3,  
can be possible via the RDI mechanism. 

Now we will discuss  the bipolar \hii region and its importance on star formation processes. In bipolar case, it is most likely that the \hii region forms in the dense region of a filamentary cloud, where the density along the equatorial axis of the filament is high, whereas it is low in the polar directions, leading to a high density contrast. As a consequence, matter can be more compressed in the equatorial plane during the expansion of the \hii region, depending upon the initial density and ionization radiation from the \hii region. A prototype example of such bipolar morphological \hii region is ``Sh2-201", where two dense ($>$ 10$^{22}$ cm$^{-2}$) and massive ($>$ 70 \msun) condensations are found at each sides of the waist of Sh2-201 (Deharveng et al. 2012). Sh2-201 is believed to be formed in a large filament that is running east-west of Sh2-201. Massive condensations are the potential 
sites of new star formation. Indeed, Deharveng et al. (2012) detected two 100 $\mu$m point sources possibly of Class 0 nature in the east condensation ($\sim$ 235 \msun) and hypothesized that the triggered  star formation is in act at the waist of Sh2-201, which
seems to be in accordance with numerical simulations of triggered star formation in a filamentary cloud due to dynamical compression of an expanding \hii 
region (see Fukuda  \& Hanawa 2000). In the absence of long wavelength observations, it 
is tempting to guess about the shape and structure of original cloud in which Sh2-294 has formed. However, the morphology of Sh2-294 at 8 $\mu$m and 22 $\mu$m on a smaller scale suggests that it might be filamentary in nature. 
Presuming Sh2-294 formed in a filamentary cloud, we compared the observed properties with the numerical simulations. 
The numerical simulations by Fukuda  \& Hanawa (2000), including a wide variety of physical conditions, suggest for typical conditions the \hii region takes more than five times the sound crossing time ($t_{cross}$: see Eq. (19)  of Fukuda  \& Hanawa 2000), to form the first generation cores (i.e., cores at the waist) in a filament. The first generation cores can be followed  by two second generation cores depending on the initial conditions due to the change in self gravity in the filament. The second generation cores 
in a magnetized cloud are expected to be separated from the first generation cores by the wavelength of most probable 
fragmentation, which is thirteen times the length scale ($H$: see Eq. (18)  of Fukuda  \& Hanawa 2000). The $t_{cross}$ and $H$ in the simulations of Fukuda  \& Hanawa (2000) depend on the sound speed and  the central density of the cold filamentary cloud. 
In Sh2-294, we presume the H$_2$ structures $\#1$, and  $\#3$, situated $\sim$ 1.7 and 1.8 pc away from the ionizing source along the 
long axis seen in 8 $\mu$m, perhaps represent the externally heated part of the condensations that are at the waist of Sh2-294.
Assuming an effective sound speed  $\sim$ 0.6 \kms, as found  in central 
region of filaments (e.g., Heitsch et al. 2009; the observed velocity dispersion)  and the average density $\sim$ 10$^5$-10$^4$ cm$^{-3}$ of the 
protocluster forming clumps (e.g., Mueller al. 2002; Motte et al. 2008), as the density of cluster forming filaments, we estimated the length scale 
and sound crossing time for Sh2-294 as $\sim$ 0.04-0.14 pc and $\sim$  7.6-25$\times$10$^4$ yr, respectively. The presence
of second generation cores cannot be assessed with the present data, but what we see today are certainly  the 
presence of two possible H$_2$ structures ($\#$1 and $\#$3), with Class I YSOs (C1 and C3) inside of these structures. 
Assuming that these Class I YSOs are resulted of the cores that might have formed due to the dynamical compression of the expanding \hii region as simulated by Fukuda  \& Hanawa (2000), we calculated the approximate time that might have elapsed in 
the entire process. In the simulations of Fukuda  \& Hanawa (2000), the time required to produce dense cores depends upon several 
factors.
We considered those models, which include magnetic field,
treat the  formation of \hii region on the filament axis or very close to it ($<$ 5$H$) as well as impinging  the filament axis with a kinetic energy. These models predict the time required to 
produce two first generation cores ($\rho$ $>$ $\sqrt10$ $\rho$$_0$) is less than or equal to ten times 
the crossing time scale, which is $\leq$ 2.5 $\times$ 10$^6$ yr in the present case. The age of the Class I YSOs is of the order of $\sim$ 10$^5$ yr. Assuming that the age ($\sim$ 4 $\times$ 10$^6$ yr) of the ionizing source represents the age of 
\hii region, it seems that Sh2-294 is old enough to produce two cores, which perhaps further 
collapses to form the Class I YSOs

 The above discussion, with the present limited observations suggests that both the hypotheses can be justified being responsible for triggered star formation in Sh2-294 region.  It is difficult to conclude with the present observations, which one is superior over the other, though we prefer triggered star formation  at the waist of Sh2-294 in a filamentary cloud, as the overall morphology of Sh2-294 on a smaller scale  resembles with a filamentary cloud. However, the proof of this would require additional informations such as determination of exact stellar properties of YSOs with infrared spectroscopic observations, a search  for in favor of possible filamentary cloud with deep FIR observations and kinematics with high resolution line observations.
\section{Summary}
In Paper-I we studied the Sh2-294 region and 
properties of the associated central cluster. With the help of low resolution 
radio continuum observation at 1280 MHz, mid \& far-infrared emissions from MSX \&
HIRES  maps and optical-NIR point source analyses, we tried to
 understand for the first time, the star formation scenario in the Sh2-294 region. 
In the present paper the high
resolution {\it Spitzer} images along with H$_2$ image with VLT
reveal the presence of several dusty structures and four prominent 
H$_2$ arc-like structures. In this paper we focus on the identification of YSOs 
and tried to constrain their physical properties using colour–-colour selections 
and SED fitting models. In  this study, we identified 25 Class II, 6 Class I and 5 Class I/II sources on the basis of {\it Spitzer}
colours and estimated their  basic parameters such as age, mass, disk accretion rate, and extinction 
using SED fitting models. These parameters provide additional insight into the evolutionary status of these YSOs, thus help us to establish a spatial and temporal correlation among the YSOs and the ionizing source of the Sh2-294. The Class I sources are found to be preferentially situated at
the periphery of the \hii region and are associated with H$_2$ arc-like structures. 
We have also shown that the population of Class I sources is indeed young in comparison to Class II YSO population and B0 MS star at the center of the  nebula. The Class I 
source  associated with the structure  ``$\#1$",  is a massive ($\sim$9 \msun) and
deeply embedded (\av~$\sim$24 mag) protostar, accreting currently with  an enveolpe accretion rate 
$\sim$ 9.8 $\times$ 10$^{-4}$ \msun~yr$^{-1}$. From the morphology, spatial distribution of YSOs and the time scale involved, 
we discussed  triggered star formation in the region with two possible scenarios.
The overall morphology of Sh2-294 along with the time scale involved suggest that the dynamical compression of a filamentary cloud by the expanding \hii region, 
following the simulations of Fukuda \& Hanawa (2000), might be the cause for new star fromation in the region. However, further observations are needed to confirm 
this possibility.
\section*{Acknowledgments}
We are grateful to an anonymous referee for his/her constructive criticisms 
that have helped us to improve the scientific content and give a new direction to the discussion on star formation activity.
MRS is grateful to  Lise Deharveng for discussion on bipolar nebula and providing a copie of her work on W5 complex with {\it Herschel} prior to publication, which includes Sh2-201 discussed in the present draft. 
This work  is based  on  observations made with  the {\it Spitzer}
Space Telescope,  which is operated by the  Jet Propulsion Laboratory,
California Institute  of Technology under a contract  with NASA.  This publication makes 
use of  the data products from  the  2MASS,  which is  a  joint  project  of the  University  of
Massachusetts    and   the    Infrared    Processing   and    Analysis
Center/California  Institute  of Technology.
This publication makes use of data products from the WISE, which is a joint project of the University of California, Los Angeles, and the Jet Propulsion Laboratory/California Institute of Technology, funded by the ``National Aeronautics and Space Administration." This work also uses the observations with AKARI, a JAXA
project with the participants of ESA.

\end{document}